\documentclass[aps,prd,preprint]{revtex4-1}
\usepackage{amssymb,graphicx,color,subfig}
\newcommand{\lsim}{\mathrel{\mathop{\kern 0pt \rlap
  {\raise.2ex\hbox{$<$}}}
  \lower.9ex\hbox{\kern-.190em $\sim$}}}
\newcommand{\gsim}{\mathrel{\mathop{\kern 0pt \rlap
  {\raise.2ex\hbox{$>$}}}
  \lower.9ex\hbox{\kern-.190em $\sim$}}}

\newcommand{\beq}{\begin{equation}}
\newcommand{\eeq}{\end{equation}}
\newcommand{\be} {\begin{eqnarray}}
\newcommand{\ee} {\end{eqnarray}}

\newcommand{\met}{\rlap{/}{\!E_T}}
\newcommand{\ie}{\textit{i.e.}, }

\newcommand{\br}{{\rm Br}}

\newcommand{\pb}{{\,{\rm pb}}}

\newcommand{\gev}{{\,{\rm GeV}}}
\newcommand{\tev}{{\,{\rm TeV}}}
\newcommand{\sg}{{\sigma}}

\newcommand{\Lm}{{\Lambda}}

\newcommand{\no}{{\nonumber}}
\newcommand{\n}{ {(n)} }

\newcommand{\tw}{ {(2)} }
\newcommand{\z}{ {(0)} }
\newcommand{\wtw}{W^{(2)}}

\begin{document}

\title{
Implications on the mUED model \\ from the early
LHC data on $\ell+\met$ signal.
}
\bigskip

\author{Sanghyeon Chang}
\email{sang.chang@gmail.com}
\author{Kang Young Lee}
\email{kylee14214@gmail.com}
\author{Jeonghyeon Song}
\email{jhsong@konkuk.ac.kr}
\affiliation{
Division of Quantum Phases \& Devices, School of Physics, 
Konkuk University,
Seoul 143-701, Korea
}
\date{\today}

\begin{abstract}
Recently the ATLAS and CMS collaborations
reported their search for a new heavy gauge boson $W'$
with one lepton plus missing transverse momentum.
We find that $W^{(2)}$,
the second Kaluza-Kelin (KK) state of the $W$ boson
in the minimal Universal Extra Dimension (mUED) model,
can be a good candidate for this signal,
as its branching ratio into $\ell\nu$ is sizable.
Moreover, nearly degenerate
KK mass spectra in the mUED model
yield generically very soft SM particle accompanying $W^{(2)}$
from the subsequent decays of the second KK quarks and gluons.
In a hadron collider,
this indirect $W^{(2)}$ production is difficult to distinguish from
the $W^{(2)}$ single production.
The involved strong interactions 
make it more important than the single production.
The early LHC data on $\ell+\met$ signal 
for 1.1 fb$^{-1}$ integrated luminosity
is shown insufficient to limit our model even with
significant indirect production of $W^{(2)}$.
However, the results show that
the present LHC 5.6 fb$^{-1}$ data can
cover most of the reasonable parameter space of the mUED model.
\end{abstract}

\maketitle

\section{Introduction}
\label{sec:introduction}

The performance of the LHC in 2010 and 2011
has been captivatingly successful.
Initial goal of integrated luminosity in 2011
was 1 fb$^{-1}$, 
but already 5.6 fb$^{-1}$ data have been delivered,
respectively, to the
ATLAS and CMS detectors 
by the end of August in 2011~\cite{LHC:prospect}.
Even with partial and early data of the LHC,
significant constraints have been made on
many new physics models 
such as supersymmetry models~\cite{early:susy},
$Z'$ models~\cite{early:zp},
and $W'$ models~\cite{early:wp}.
   
One of the most sensitive 
and clean probe for new physics 
is the event with a highly energetic electron or muon 
and the large missing transverse energy $\met$.
The CMS~\cite{cms1} and ATLAS collaborations~\cite{atlas1}
have reported the analysis of $\ell+\met$ data
corresponding to an integrated luminosity of 36 pb$^{-1}$.
Both experiments found
no excess beyond the SM expectations.
Using a reference $W'$ model,
in which a heavy $W'$
has the same left-handed fermionic couplings
and vanishing interactions with the SM gauge bosons,
the lower bound on the $W'$ mass 
has been made to be about 1.4 TeV.
Recently it is far more refined to be 2.27 TeV
with 1 fb$^{-1}$ luminosity data collected in 2011~\cite{2011data}.
Recently their implications on various new physics models,
such as non-universal gauge interaction model~\cite{KY:non-universal},
minimal walking technicolor model~\cite{mTC}, and
left-right model~\cite{LR},
have been extensively studied.

We find that the Universal Extra Dimension 
(UED) model has a good candidate to mimic the $W'$
decaying into $\ell\nu$,
the second Kaluza-Klein (KK) mode of the $W$ boson, $\wtw$.
In addition, the minimal version of the UED model, 
called the mUED model, has 
additional enhancement of the $\wtw$ production at the LHC.
The UED model is 
based on a single flat extra dimension of size $R$,
compactified on an $S_1/Z_2$ orbifold.
This fifth dimensional space is accessed by all the SM fields.
Thus all the SM fields have an infinite number of KK 
excited states
of which the zero modes are identified to the SM fields.
At tree level
the KK number $n$ is conserved
by the fifth dimensional momentum conservation,
but broken to the KK parity at loop level.
Due to the KK parity conservation,
the lightest KK particle (LKP) with odd KK parity is stable
and becomes a good candidate of the cold dark matter. 
In the mUED model, 
the boundary kinetic terms are assumed to vanish 
at the cut-off scale $\Lm$.
Radiative corrections to the KK masses 
are finite and calculable:
the first KK mode of the $U(1)_Y$ gauge boson 
$B^{(1)}$ is the LKP~\cite{rad-correction}. 
The thermal relic density of $B^{(1)}$
with mass around 500 GeV can explain 
all of the dark matter~\cite{KK:DM}.
In order to avoid over-closing the universe, 
the $B^{(1)}$ mass is constrained to be
below about $600$ GeV~\cite{DM}. 

Various phenomenological study of the mUED has been done
in the literature~\cite{pheno}.
There are two distinctive features 
that differentiate the mUED model from
other new physics models:
the nearly degenerate KK mass spectra of new particles 
and the presence of heavy parity-even ($n=2$) particles~\cite{discrimination,KKHiggs}.
These two features leave very interesting phenomenology
associated with the second KK modes, especially $\wtw$.
%
%

In this paper, 
we examine the production of the $W^{(2)}$ boson 
followed by its decay into $\ell\nu$ in the mUED model,
and study the constraints by the early LHC data.
Due to the kinematic suppression by nearly degenerate KK mass spectra,
the KK-number conserving decays of 
$W^{(2)} \to f^{(2)} \bar{f}'^{(0)}$
and $W^{(2)} \to f^{(1)} \bar{f}'^{(1)}$
are not dominant.
The KK-number violating decays into two SM fermions
at one loop level are considerable.
Moreover we notify that there are sizable indirect production 
of $W^{(2)}$ boson
in the decays of heavier colored KK states of $n=2$,
\ie the second KK quarks $Q^\tw$ and gluons $g^\tw$.
The SM particles,
which are by-products of these cascade decays, 
are generically very soft 
due to the degenerate masses of the second KK states.
Thus the transverse mass distribution of the leptons
from indirectly produced $W^{(2)}$
is similar to that from singly produced $W^{(2)}$.
This indirect production is
more important than the direct production.
This is our main result.

The paper is organized as follows.
In the section \ref{sec:model}, 
we briefly describe the model and 
discuss the production and decay of the $W^{(2)}$ boson. 
Section \ref{sec:collider} is devoted to 
the analysis with the data collected at the LHC and Tevatron.
We conclude in Sec.~\ref{sec:conclusion}.

\section{Productions and Decays of $W^{(2)}$ boson
in the mUED model}
\label{sec:model}

The UED model is based on an additional extra dimension
$y$ with size $R$ where all the SM fields propagate.
The fifth dimension $y$ is compactified on an
$S_1/Z_2$ orbifold
for generating zero mode chiral fermions.
We assign odd parity under the $Z_2$ orbifold symmetry
to the zero mode fermion with wrong chirality.
This extends the fermion sector
into $SU(2)$-doublet quark $Q(x,y)$
and $SU(2)$-singlet quark $q(x,y)|_{q=u,d}$.
After compactification, we obtain four-dimensional
effective Lagrangian with the zero modes and the KK excited states.
Focused on the phenomenology of the second KK modes of $W$ boson,
we present the relevant KK expansions of
\be
V_\mu(x,y) &=& \frac{1}{\sqrt{\pi R}}
     \left[
      V^{(0)}_{\mu}(x) + \sqrt{2} \sum_{n=1}^\infty V^{(n)}_\mu (x) 
                     \cos \frac{n y}{R}
     \right],
\\ \nonumber
V_5(x,y) &=& \sqrt{\frac{2}{\pi R}}
            \sum_{n=1}^\infty V^{(n)}_5 (x) \sin \frac{n y}{R},
\\ \no
Q(x,y) &=&
\frac{1}{\sqrt{\pi R}}
\left[
Q^\z_{L}(x) + \sqrt{2} \sum_{n=1}^\infty 
\left\{
Q_{L}^\n (x) \cos \frac{n y}{R}
+ 
Q_{R}^\n (x) \sin \frac{n y}{R}
\right\}
\right],
\\ \no
q(x,y) &=&
\frac{1}{\sqrt{\pi R}}
\left[
q^\z_R(x) + \sqrt{2} \sum_{n=1}^\infty 
\left\{
q_{R}^\n (x) \cos \frac{n y}{R}
+ 
q_L^\n (x) \sin \frac{n y}{R}
\right\}
\right],
\ee
where $V^M =B^M, W^M, A^M, g^M$, and $n$ is the KK number. 

The $n$-th KK mass of a gauge boson $V$ is given by
\be
\label{eq:KKmass}
M^2_{V^{(n)}} =  M_n^2 +  m_0^2 + \delta m^2_{V^{(n)}},
\ee
where $M_n=n/R$, $m_0$ is the corresponding SM particle mass
and $\delta m^2_{(n)}$ is the radiative corrections.
There are two types of radiative corrections
to the KK mass, which break generically five-dimensional  
Lorentz invariance.
The first is the bulk correction from compactification
or non-local loop-diagrams around the circle of the compactified
dimension $y$.
Since this propagation is over finite distances,
these bulk corrections are well-defined 
and finite.
The second type of radiative corrections
are from the boundary kinetic terms,
which are incalculable due to unknown physics 
at the cutoff scale $\Lm$.
The mUED model is based
on the assumption that the boundary kinetic terms vanish 
at the cutoff scale $\Lm$.
The radiative correction to the KK mass of the $W^{(n)}$ bosons 
is given by~\cite{rad-correction}
\be
\delta m^2_{W^{(n)}} 
= 
-\frac{5}{2} \frac{g^2 \zeta(3)}{16 \pi^4 } \frac{1}{R^2}
 + M_n^2
\frac{15}{2} \frac{g^2}{16 \pi^2}\ln\,\frac{\Lambda^2}{\mu^2}
,
\ee
where the renormalization scale $\mu$
is normally set to be $M_n$.

\begin{figure}[t!]
\centering
\includegraphics[width=6cm]{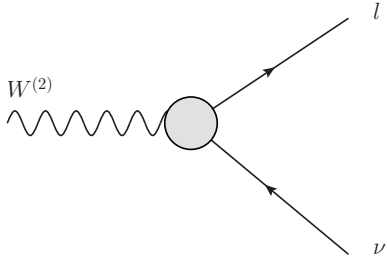}
\caption{\label{fig:Feyn:Wlnu}\small
Feynman diagrams for the  decay of the $W^{(2)}$ boson. 
}
\end{figure}

Search for a charged heavy gauge boson $W'$ at the LHC
is being conducted in its leptonic decay channels
with electron and muon final states:
see Fig.~\ref{fig:Feyn:Wlnu}.
The relevant KK-number violating operator is
\be
{\cal L}_{200} = i \hat{g}_{ff'} 
\left(\frac{g}{2}
\frac{1}{16\pi^2} \ln \frac{\Lm^2}{\mu^2}
\right)
\bar{f}  \gamma^\mu P_L f'  W_{\mu}^{(2) },
\ee
where 
\be
\hat{g}_{\ell\nu} = \frac{9}{8} g^{\prime 2} -\frac{33}{8}g^2,
\quad
\hat{g}_{q q'} = \frac{1}{8} g^{\prime 2}
- \frac{33}{8} g^2 + 6 g_s^2.
\ee
The branching ratios of $W^{(2)}$ 
have been computed in Ref.~\cite{datta}.
Depending on $R^{-1}$, ${\rm Br}(W^{(2)} \to l \nu) \sim 2-3 \%$.

\begin{figure}[t]
\centering
\includegraphics[width=6cm]{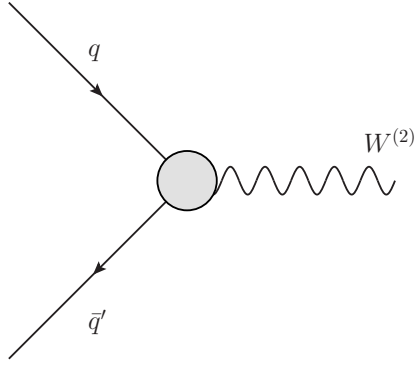}
\caption{\label{fig:single:W2}\small
Feynman diagrams for the single production of the $W^{(2)}$ boson. 
}
\end{figure}

As depicted in Fig.~\ref{fig:single:W2},
the single production of $W^{(2)}$ boson is
through the KK-number violating operator ${\cal L}_{200}$
with $f=q$.
The production cross section
is $\sigma(p p \to W^{(2)}) \sim {\cal O}(0.1) \pb$
for $1/R=500$ GeV~\cite{datta}.

\begin{figure}[t]
\centering
\subfloat[]
{ \includegraphics[width=6cm]{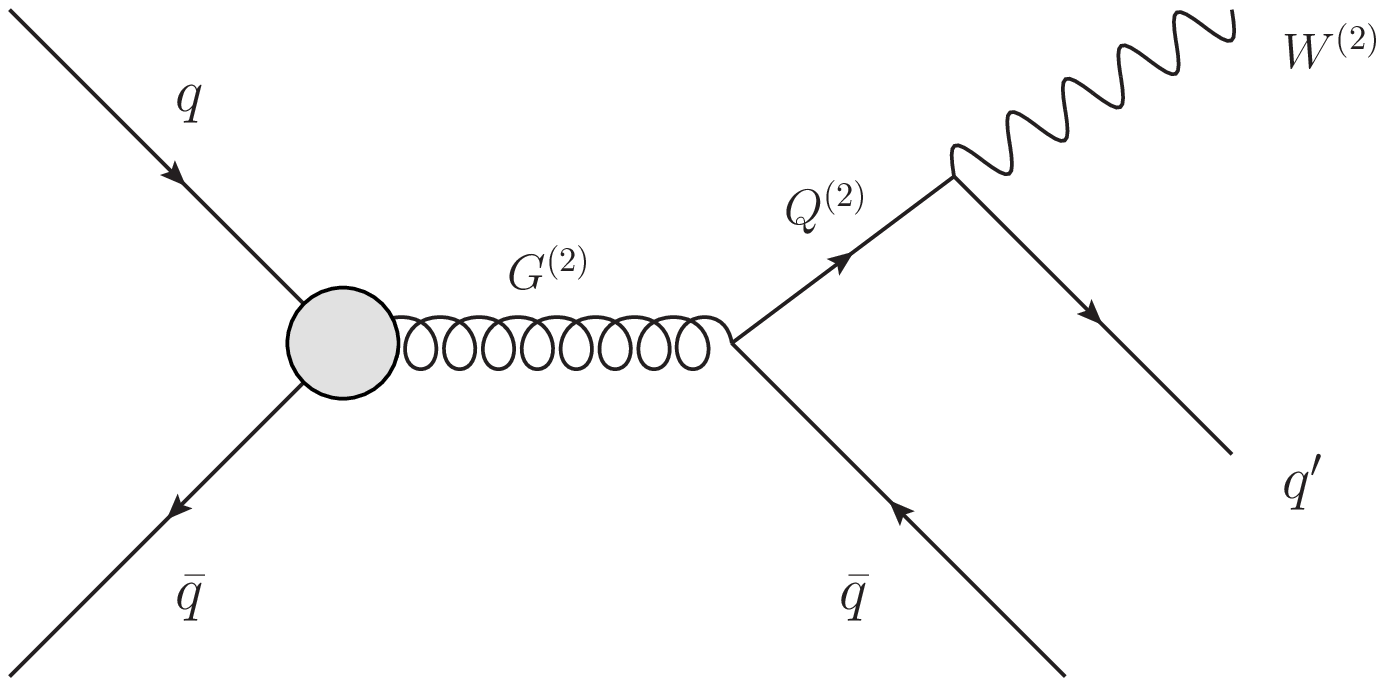} }
\subfloat[]
{ \includegraphics[width=6cm]{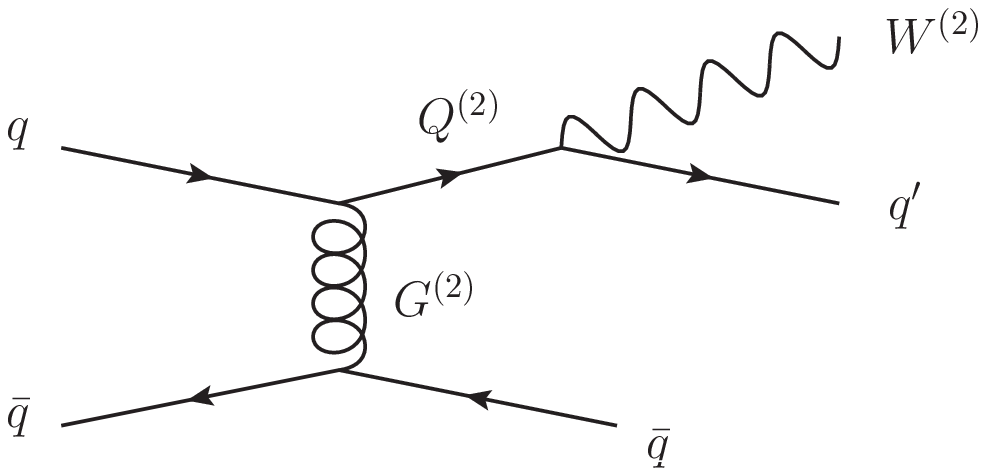} }
\\
\subfloat[]
{ \includegraphics[width=6cm]{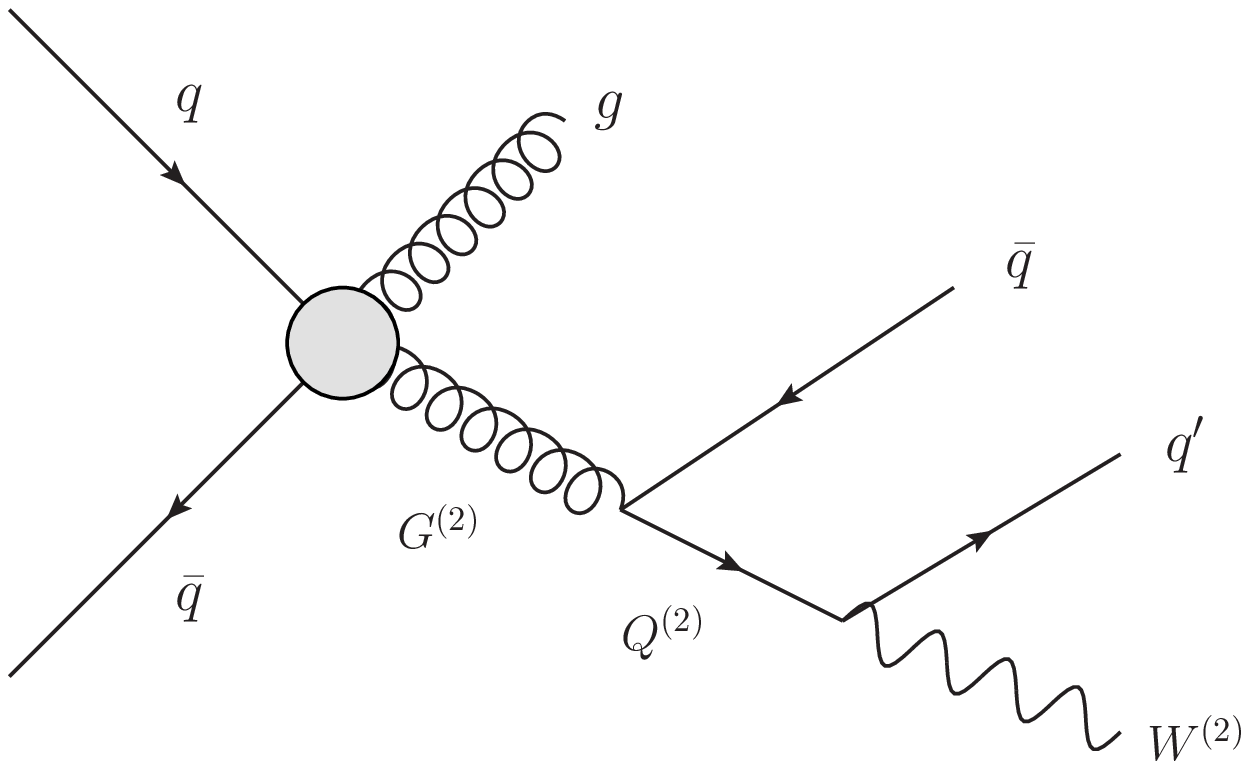} }
\subfloat[]
{ \includegraphics[width=6cm]{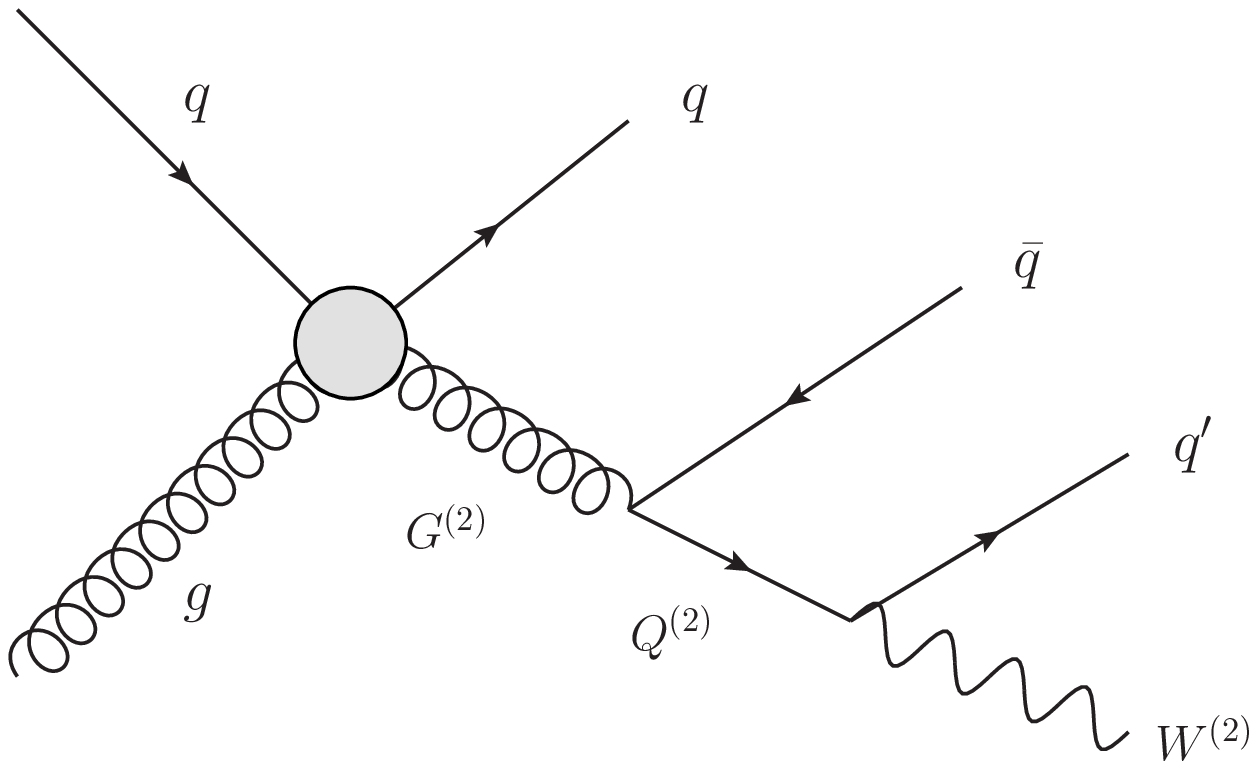} }
\caption{\label{fig:indirect:W2}\small
Feynman diagrams for the production of the $W^{(2)}$ boson. 
}
\end{figure}

At the LHC, the $W^{(2)}$ boson is also produced 
through the cascade decays 
of a heavier second KK modes, $Q^\tw$ and $g^\tw$.
In the mUED model,
the KK mass spectra are unambiguously fixed, 
leading to the
hierarchy of 
$M_{g^\tw} > M_{Q^\tw} > m_{W^\tw}$.
As shown in Fig.~\ref{fig:indirect:W2},
the second KK gluon $g^\tw$
can decay into $Q^\tw q$ 
with branching ratio of about 50\%,
and $Q^\tw$ decays into $\wtw q'$ with branching ratio of about 50\%.
Small mass differences of $M_{g^\tw} - M_{Q^\tw}$
and $M_{Q^\tw}-M_{\wtw}$
make the accompanying SM quarks very soft.
At a hadron collider,
the phenomenological signature 
of the indirectly produced $W^{(2)}$ boson
is likely to be indistinguishable 
from that of the singly produced $W^{(2)}$.
As shall be shown, 
this indirect production of $\wtw$ is more important.

Figure \ref{fig:indirect:W2} illustrates 
the indirect production of $\wtw$
accompanying soft jets.
In Fig.~\ref{fig:indirect:W2}(a),
$g^\tw$ is singly produced, followed by its decay of  
$g^\tw \to Q^\tw q$ and $Q^\tw \to \wtw q'$.
Nearly degenerate mass spectrum yields
very soft jets.
Note that the single production of $Q^\tw$
is not possible since the leading vertex $g$-$q$-$Q^\tw$ 
vanishes as required by gauge invariance~\cite{rad-correction}.
Figures \ref{fig:indirect:W2}(b)-(d)
present associated production of the heavy second KK mode
with a SM quark or gluon,
$pp \to \bar{q} Q^\tw ,~  g g^\tw, ~q  g^\tw $.
In order to show the softness of the accompanying SM jet,
we present the four-momenta of the heavy 
second KK mode and the SM particle
in the parton c.m frame:
\be
k^\mu_{\tw}
=
\left(
\sqrt{E^2 + M_\tw^2 }, E
\right),
\quad
k^\mu_j
=\left(
E, -E
\right),
\hbox{ where } E = \frac{\hat{s} - M_\tw^2 }{2 \sqrt{\hat{s}}}.
\ee
The steeply falling parton luminosities
lead to the production of new heavy particles 
near the threshold at the LHC:
the energy of the accompanying SM particle is
quite low.

We summarize the indirect production processes as
\be
p p & \to & q \bar{q} \to Q^{(2)} \bar{q} \to W^{(2)} q' \bar{q},
\nonumber \\
p p & \to & q \bar{q} \to G^{(2)} g \to Q^{(2)} \bar{q} g 
        \to W^{(2)} q' \bar{q} g,
\nonumber \\
p p & \to & g q \to G^{(2)} q \to Q^{(2)} \bar{q} q \to W^{(2)} q' \bar{q} g.
\ee
We impose the condition of the soft SM particles as 
$1 \gev < p_T^j < 30\gev$ at the LHC. 
Note that the lower cut of 1 GeV is assigned to avoid the infrared 
and collinear divergences.
Since the second KK quarks and gluons are produced 
through strong interactions,
their production rates are very high and
the number of the $W^{(2)}$ boson produced from their decays 
is considerable.


We also include the sub-leading processes of the $W^{(2)}$ 
productions associated with a quark or a gluon,
\ie
$p p \to q \bar{q} \to \wtw g$ and 
$p p \to g q \to \wtw q$.
At the LHC, the cross section of $p p \to  \wtw g$
is much larger than that of $p p \to \wtw q$.
The same soft $p_T^j$ cut is applied.

\section{Implications on the $W^{(2)}$ mass with the early LHC data}
\label{sec:collider}

The CMS and ATLAS collaborations have reported 
the results of search for $W'$ boson through the leptonic decay channel.
These events are triggered by a single isolated 
high-$p_T$ lepton
and the missing transverse energy 
of opposite direction and similar in magnitude.
The transverse mass of the $W'$ boson for candidate events
is calculated as
$M_T = \sqrt{2 p_T \met (1- \cos\phi)}$,
where $\phi$ is the azimuthal
opening angle between the lepton and the $\met$.
From the absence of the signal events 
in the early LHC data corresponding to
an integrated luminosity of 1.1 fb$^{-1}$,
an upper limit at 95\% C.L.
on the production cross section of $W'$ 
times the branching ratio of its decay into $\ell\nu$
is set as a function of its mass.
The present bounds on the mass of $W'$
is in a reference model with SM couplings:
at 95\% C.L. the mass bounds are
2.27 TeV (CMS)
with 1.03 fb$^{-1}$ electron data and 1.13 fb$^{-1}$ muon data,
and 2.23 TeV (ATLAS) with 1.04 fb$^{-1}$ data.

\begin{figure}[t]
\centering
\includegraphics[width=12cm]{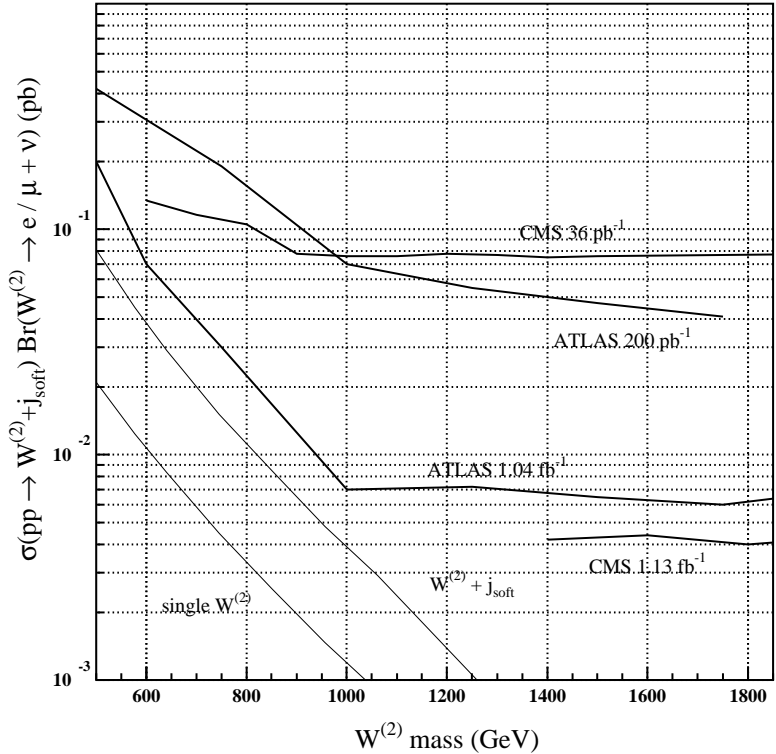}
\caption{\label{fig:LHC}\small
LHC limits with a counting experiment in the search window
$pp \to W' \to e \nu /\mu \nu$ for the $W^{(2)}$ boson
in the mUED model.
}
\end{figure}

By comparing $\sg( pp \to \wtw j_{\rm soft})
\times \br(\wtw \to \ell\nu)$
with the experimental upper limit,
we can determine the lower limit of the $W^{(2)}$ mass,
which is shown in Fig.~\ref{fig:LHC}
with the reported CMS and ATLAS data.
In order to show the importance of the indirect production of $\wtw$,
we separately present the events only from the single production
and those including indirect production with soft jets.
It is clear that the indirect production
of $\wtw$ is more important than
the signal production of $\wtw$.
For example, 
$M_{\wtw}=600\gev$ case
has the indirect production
of $\wtw$ larger than
its single production by a factor of four.
Even with enhanced production from the indirect production
of $\wtw$, however,
the current upper limit is not sufficient
to give a significant constraint on the $W^{(2)}$ mass.

We still remain optimistic about the future
prospect  of the LHC on probing the mUED model
through the $\wtw\to\ell\nu$ channel.
As can be seen in Fig.~\ref{fig:LHC},
the enhancement of the integrated luminosity,
from 200 pb$^{-1}$ to 1.1 fb$^{-1}$ at the ATLAS,
improves the sensitivity on the upper limit on the $W'$ mass
by the factor of almost ten.
With the current 5.6 fb$^{-1}$ data,
the ATLAS and CMS are very likely to probe the mUED model
for $M_{\wtw} \lsim 1\tev$.
By the end of 2012, 
we expect at least 10 fb$^{-1}$ data per experiment at the LHC.
In the near future,
the $\wtw\to\ell\nu$ channel 
is to cover most parameter space 
$M_{\wtw} \lsim 1.2 \tev$, which is allowed 
by the observed relic density.
And the inclusion of indirect $\wtw$ production 
is crucial for the future prospect.

\begin{figure}[t]
\centering
\includegraphics[width=12cm]{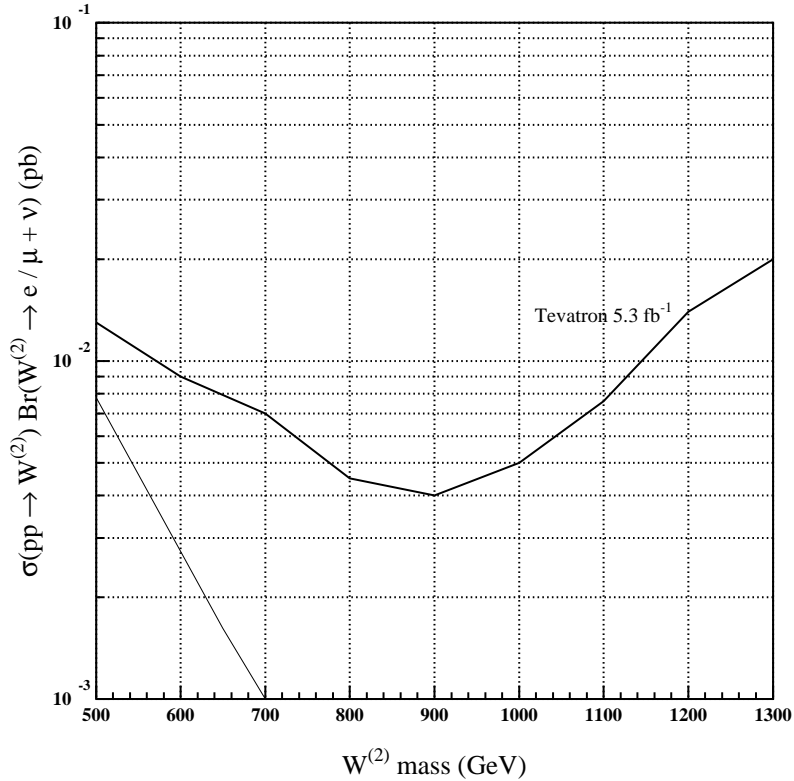}
\caption{\label{fig:Tevatron}\small
Limits with the Tevatron data in the search window
$pp \to W' \to e \nu/\mu \nu$ for the $W^{(2)}$ boson
in the mUED model.
}
\end{figure}

For completeness, we present the Tevatron limit with
the data of 5.3 fb$^{-1}$  
in Fig.~\ref{fig:Tevatron}.
The $p_T$ cut on the accompanying soft jet
is $ 1\gev < p_T < 20\gev$.
We find that the Tevatron data cannot give any constraint
on the mUED model either.
Even with the full data corresponding to an 
integrated luminosity of 10 fb$^{-1}$,
it is difficult to probe the mUED model through this channel.

\section{Concluding Remarks}
\label{sec:conclusion}

We have studied the production of the $W^{(2)}$ boson 
at the LHC to obtain the direct bound on the mUED model. 
We find that including indirect productions of $W^{(2)}$ boson
increases the production cross section by a few times and
much improve the sensitivity of the bound on the $W^{(2)}$ mass.
The reported LHC analysis
based on the data corresponding to an integrated luminosity
1 fb$^{-1}$ is not sufficient to put
the direct bound.
However, we expect that 
the currently accumulated data of 5.6 fb$^{-1}$
will yield significant limit on the mUED model.
It would be the first direct bound of the mUED model.

\acknowledgments
This work is supported by WCU program through the KOSEF funded
by the MEST (R31-2008-000-10057-0).
KYL is also supported by
the Basic Science Research Program 
through the National Research Foundation of Korea (NRF) 
funded by the Korean Ministry of
Education, Science and Technology (2010-0010916) and.
SC is supported  by the Basic Science Research Program 
through the NRF funded by the Korean Ministry of
Education, Science and Technology  (KRF-2008-359-C00011).

\def\PRD #1 #2 #3 {Phys. Rev. D {\bf#1},\ #2 (#3)}
\def\PRL #1 #2 #3 {Phys. Rev. Lett. {\bf#1},\ #2 (#3)}
\def\PLB #1 #2 #3 {Phys. Lett. B {\bf#1},\ #2 (#3)}
\def\NPB #1 #2 #3 {Nucl. Phys. B {\bf #1},\ #2 (#3)}
\def\ZPC #1 #2 #3 {Z. Phys. C {\bf#1},\ #2 (#3)}
\def\EPJ #1 #2 #3 {Euro. Phys. J. C {\bf#1},\ #2 (#3)}
\def\JHEP #1 #2 #3 {JHEP {\bf#1},\ #2 (#3)}
\def\IJMP #1 #2 #3 {Int. J. Mod. Phys. A {\bf#1},\ #2 (#3)}
\def\MPL #1 #2 #3 {Mod. Phys. Lett. A {\bf#1},\ #2 (#3)}
\def\PTP #1 #2 #3 {Prog. Theor. Phys. {\bf#1},\ #2 (#3)}
\def\PR #1 #2 #3 {Phys. Rep. {\bf#1},\ #2 (#3)}
\def\RMP #1 #2 #3 {Rev. Mod. Phys. {\bf#1},\ #2 (#3)}
\def\PRold #1 #2 #3 {Phys. Rev. {\bf#1},\ #2 (#3)}
\def\IBID #1 #2 #3 {{\it ibid.} {\bf#1},\ #2 (#3)}

\end{document}